\documentclass[a4paper,11pt]{article}
\pdfoutput=1

\usepackage{jheppub}
\usepackage[T1]{fontenc}
\usepackage[caption=false]{subfig}
\usepackage{booktabs}
\usepackage{makecell}
\hypersetup{hypertexnames=false}

\title{Search for visible dark photon decays at the future SHINE facility with the DarkSHINE experiment}

\author[a]{Zejia Lu,}
\author[b,c]{Huayang Wang,}
\author[b]{Xiang Chen,}
\author[b]{Jiahui Wu,}
\author[d]{Yulei Zhang,}
\author[b,1]{and Liang Li\note{Corresponding author.}}

\affiliation[a]{State Key Laboratory of Dark Matter Physics, School of Physics and Astronomy and Tsung-Dao Lee Institute, Key Laboratory for Particle Astrophysics and Cosmology (MOE), Shanghai Key Laboratory for Particle Physics and Cosmology, Shanghai Jiao Tong University,\\Shanghai, China}
\affiliation[b]{State Key Laboratory of Dark Matter Physics, School of Physics and Astronomy, Key Laboratory for Particle Astrophysics and Cosmology (MOE), Shanghai Key Laboratory for Particle Physics and Cosmology, Shanghai Jiao Tong University,\\Shanghai, China}
\affiliation[c]{Shanghai Innovation Institute,\\Shanghai, China}
\affiliation[d]{Department of Physics, University of Washington,\\Seattle, Washington, USA}

\emailAdd{liangliphy@sjtu.edu.cn}

\abstract{
We study the sensitivity of the proposed DarkSHINE experiment to displaced dark photon decays, $A'\to e^+e^-$, using an 8~GeV electron beam incident on a tungsten target. We consider $2m_e<m_{A'}<2m_\mu$ and assume a unit branching fraction to electron--positron pairs. Signal events generated at tree level with \textsc{CalcHEP} are propagated through a detailed \textsc{Geant4} detector simulation. A close tracker, graph-neural-network-based track reconstruction, and displaced vertex selections suppress prompt pair-production and photon-conversion backgrounds, while calorimeter and invariant-mass requirements render hadronic backgrounds negligible. Extrapolation of the residual prompt-background distribution yields 0.3 expected events for $9\times10^{14}$ electrons on target. The median expected 90\% confidence-level exclusion covers masses of approximately 30--70~MeV and kinetic mixing parameters from approximately $5\times10^{-5}$ to $2\times10^{-4}$. This visible decay search complements DarkSHINE searches for invisible dark photon decays.
}


\begin{document}

\maketitle
\flushbottom

\section{Introduction}
Dark matter below the GeV scale has attracted substantial interest because conventional nuclear recoil experiments rapidly lose sensitivity in this mass range. This limitation has motivated accelerator searches at the intensity frontier, including experiments with fixed targets~\cite{Battaglieri:2017aum,Graham:2021ggy}. A simple and widely studied framework is the vector portal, in which a new $U(1)_D$ gauge boson, the dark photon ($A'$), couples to the Standard Model through kinetic mixing~\cite{Holdom:1985ag,Fabbrichesi:2020wbt}. The dark photon provides a benchmark for light dark matter production and broader searches for hidden sectors~\cite{Battaglieri:2017aum,Graham:2021ggy}.

DarkSHINE is a proposed experiment at the Shanghai High Repetition Rate XFEL and Extreme Light Facility (SHINE) that will use an electron beam at the GeV scale~\cite{chen2023prospective}. Depending on the dark sector spectrum, dark photons may decay invisibly into dark matter or visibly into Standard Model particles~\cite{Fabbrichesi:2020wbt}. Although invisible decays are a primary target of light dark matter searches, visible decays probe scenarios in which Standard Model final states are sizable or dominant. At small kinetic mixing, a dark photon can travel a measurable distance before decaying and produce a displaced vertex with strong rejection of prompt backgrounds~\cite{Graham:2021ggy}. Building on the projected DarkSHINE sensitivity to invisible decays~\cite{chen2023prospective}, this work studies the visible channel with a displaced vertex analysis.

Section~\ref{sec:signal_and_bkg} describes the signal and background processes. Section~\ref{sec:detector} presents the beam and detector configuration. Sections~\ref{sec:sim_and_recon} and~\ref{sec:analysis} detail the simulation, reconstruction, and analysis. Section~\ref{sec:conclusion} summarizes the results.
\section{Signal and Background Processes}
\label{sec:signal_and_bkg}

The dark photon is a $U(1)_D$ gauge boson associated with a hidden sector~\cite{Holdom:1985ag,Fabbrichesi:2020wbt}. It interacts with the Standard Model through kinetic mixing with the hypercharge field,
\begin{equation}
\mathcal{L}
= \mathcal{L}_{\mathrm{SM}}
- \frac{\epsilon_Y}{2} F^{Y,\mu\nu} F'_{\mu\nu}
- \frac{1}{4} F'^{\mu\nu} F'_{\mu\nu}
+ \frac{1}{2} m_{A'}^{2} A'^{\mu} A'_{\mu},
\end{equation}
where $\epsilon_Y$ is the kinetic mixing parameter and $m_{A'}$ is the dark photon mass. After electroweak symmetry breaking, the dark photon couples effectively to the electromagnetic current~\cite{Holdom:1985ag,Fabbrichesi:2020wbt},
\begin{equation}
\mathcal{L}_{\mathrm{int}}
= \epsilon e A'_{\mu} J^{\mu}_{\mathrm{em}},
\end{equation}
where $\epsilon = \epsilon_Y \cos\theta_W$. Dark photons can therefore be produced through bremsstrahlung in electron interactions with a fixed target~\cite{Bjorken:2009mm},
\begin{equation}
e^- + Z \rightarrow e^- + Z + A'.
\end{equation}
Using the full matrix element implemented in \textsc{CalcHEP}~\cite{Belyaev:2012qa}, we calculate the production cross section and event kinematics at tree level. The elastic nuclear form factor accounts for atomic screening and the finite nuclear size~\cite{Bjorken:2009mm,Tsai:1974fa}:
\begin{equation}
G_2^{\mathrm{el}}(t)=
\left(\frac{a^2t}{1+a^2t}\right)^2
\left(\frac{1}{1+t/d}\right)^2 Z^2,
\end{equation}
where $t$ is the squared momentum transfer, $Z$ and $A$ are the atomic and mass numbers of the target nucleus, respectively, $a=111\,Z^{-1/3}/m_e$, and $d=0.164\,\mathrm{GeV}^2 A^{-2/3}$.

\begin{figure}[htbp]
    \centering
    \includegraphics[width=0.65\linewidth]{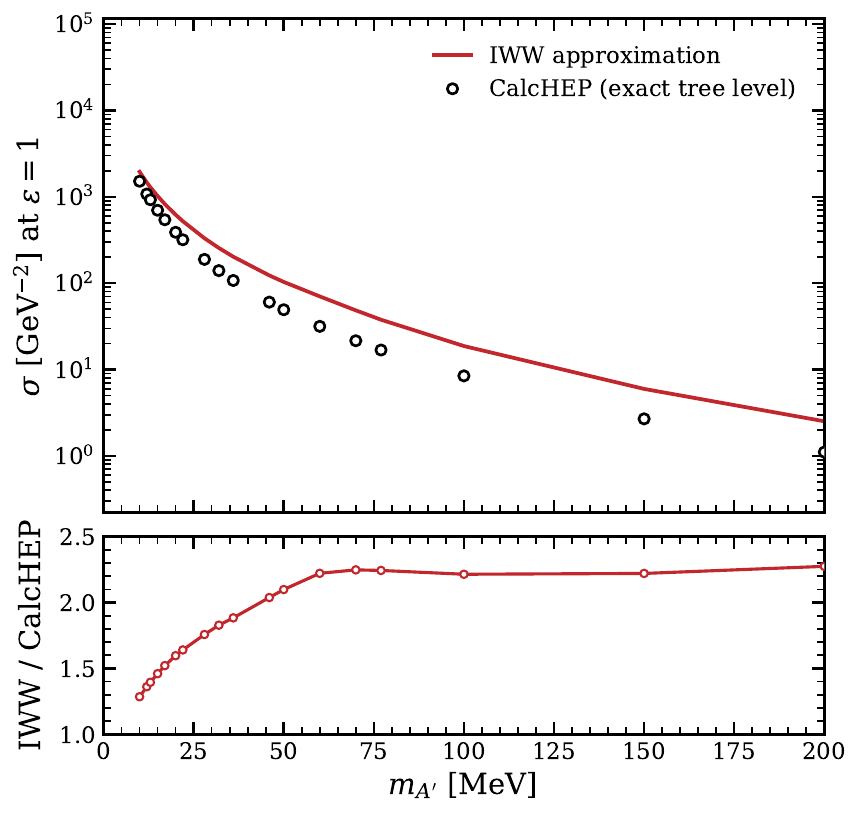}
    \caption{Dark photon production cross sections for an 8~GeV electron beam incident on a tungsten target, normalized to $\epsilon=1$. The upper panel compares the IWW approximation (red curve) with the full tree-level \textnormal{\textsc{CalcHEP}} calculation (open black circles). The lower panel shows the ratio of the IWW cross section to the \textnormal{\textsc{CalcHEP}} result.}
    \label{fig:xsec_comparison}
\end{figure}

We use the full tree-level matrix element rather than the improved Weizs\"acker--Williams (IWW) approximation~\cite{Bjorken:2009mm}, which can yield sizable deviations from the exact calculation of boson production in electron fixed-target experiments~\cite{Liu:2017ww}. Figure~\ref{fig:xsec_comparison} compares the two calculations for the DarkSHINE beam and target configuration. The IWW approximation gives a larger production cross section throughout the mass range shown. The sensitivity projections in this study use the \textsc{CalcHEP} cross sections.

Depending on the mass spectrum, the dark photon can decay into dark sector particles or visible Standard Model states~\cite{Fabbrichesi:2020wbt}. We consider $2m_e<m_{A'}<2m_{\mu}$, assuming only visible decays, and therefore $\mathrm{Br}(A'\to e^+e^-)=1$. Under this assumption, the total decay width equals the electron and positron partial width, and the inverse lifetime in natural units is~\cite{Bjorken:2009mm,Fabbrichesi:2020wbt}
\begin{equation}
\frac{1}{\tau_{A'}}
=
\Gamma(A'\to e^+e^-)
=
\frac{1}{3}\,\alpha\,\epsilon^2\,m_{A'}
\left(1+\frac{2m_e^2}{m_{A'}^2}\right)
\sqrt{1-\frac{4m_e^2}{m_{A'}^2}}.
\end{equation}

\begin{figure}[tbp]
    \centering
    \subfloat[Signal process]{%
        \includegraphics[width=0.28\textwidth]{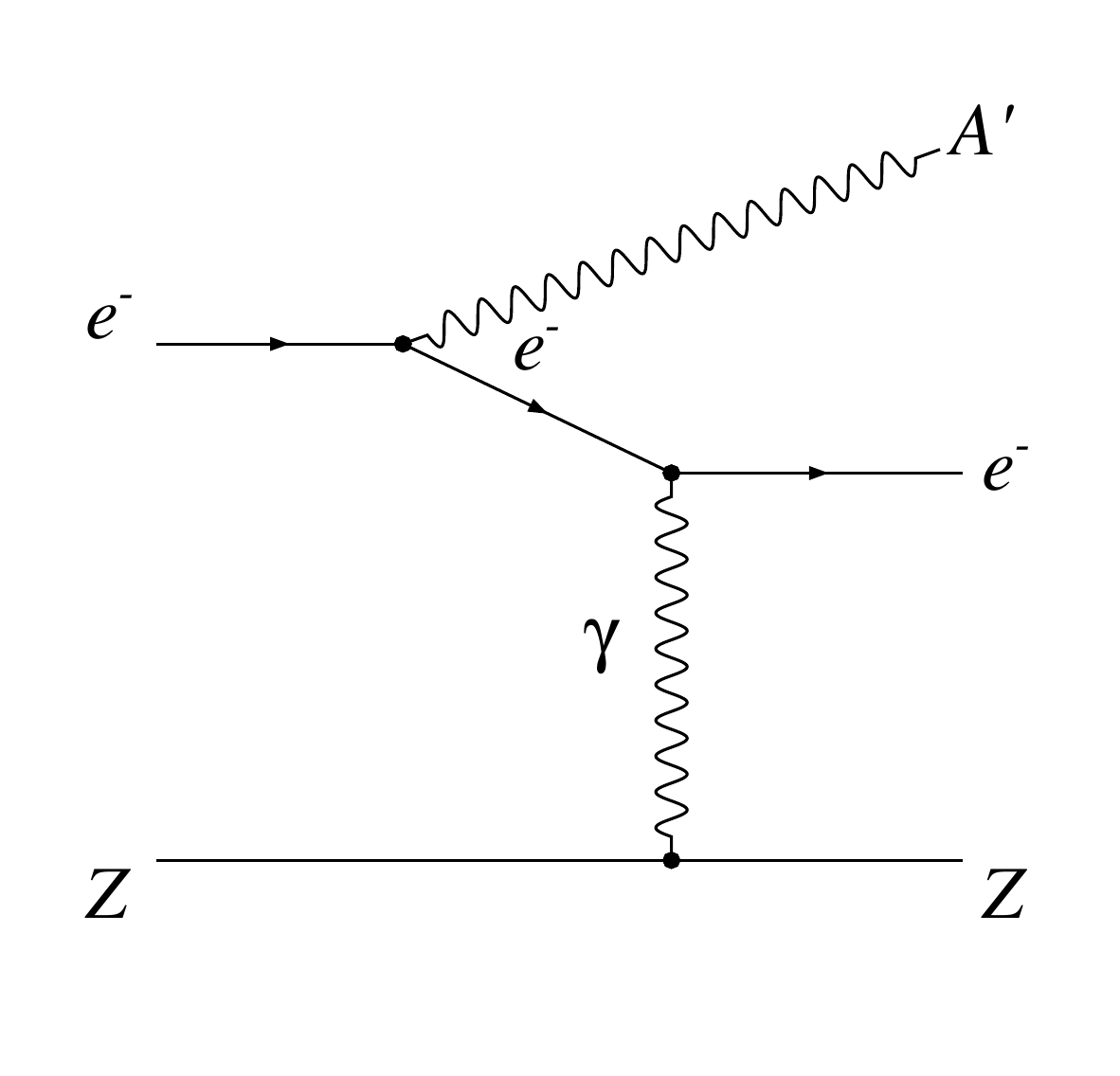}
        \includegraphics[width=0.28\textwidth]{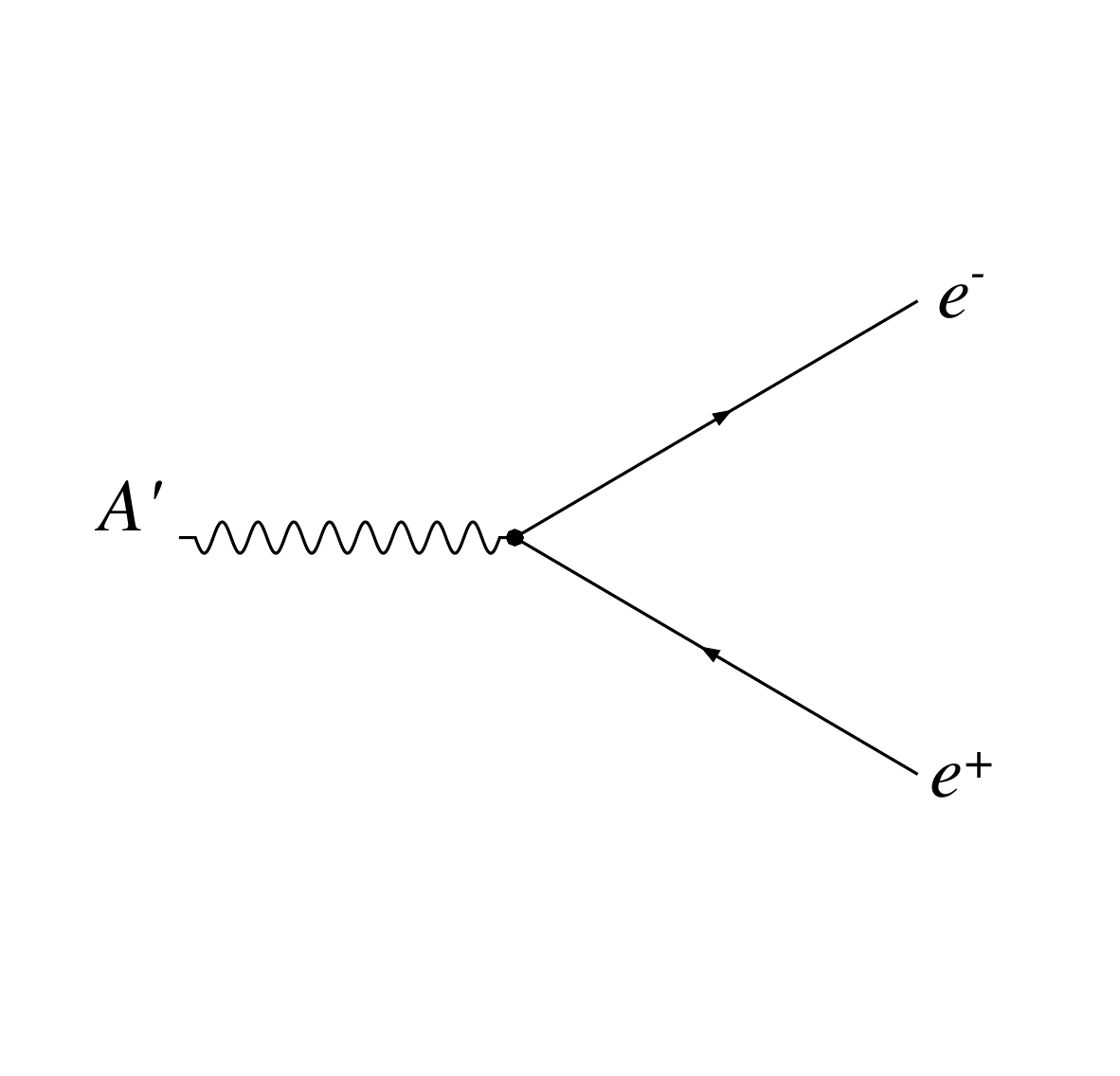}}
    \hfill
    \subfloat[Background process]{%
        \includegraphics[width=0.28\textwidth]{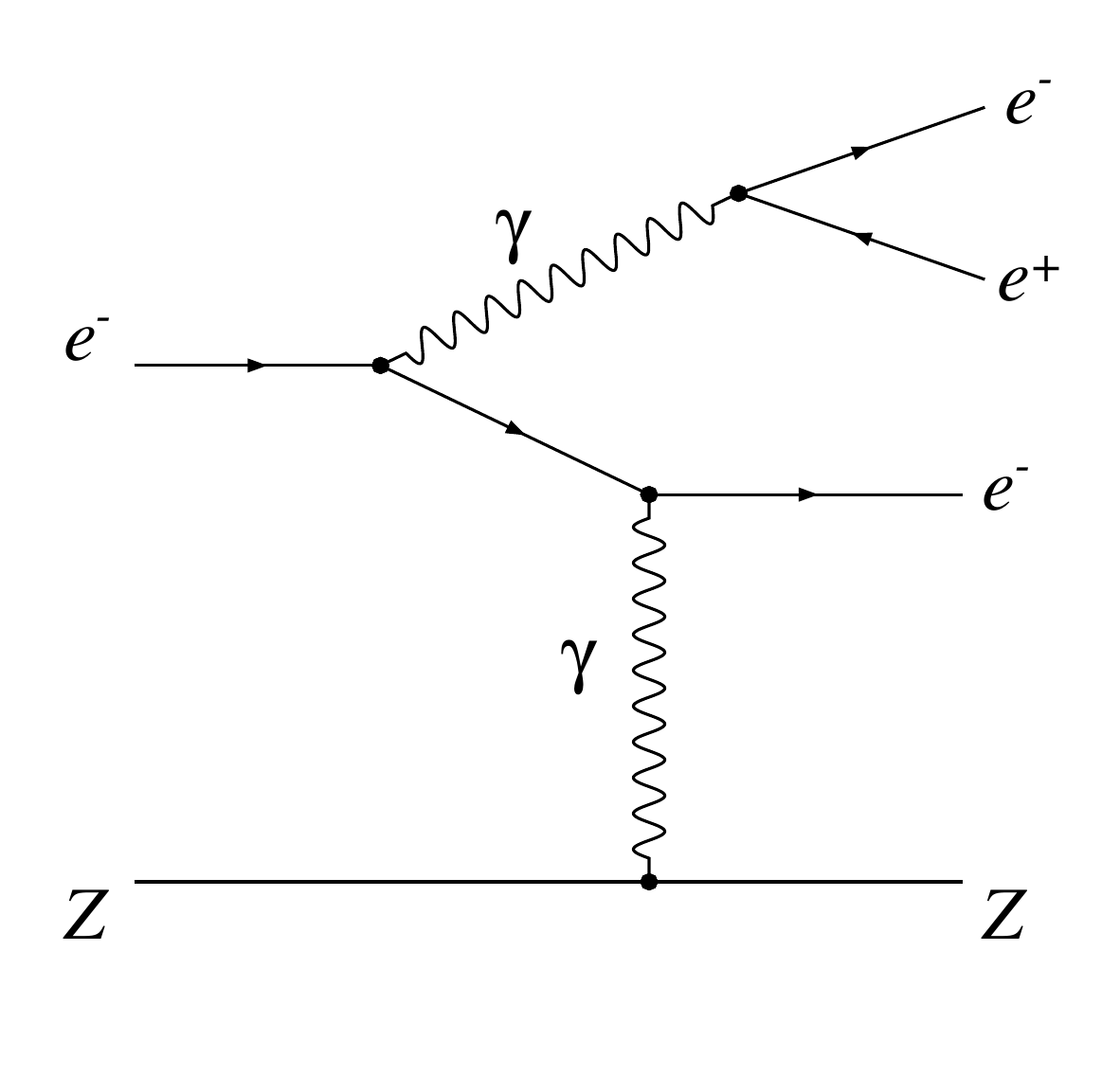}}
    \caption{Feynman diagrams for dark photon production and visible decay (left) and prompt electron and positron pair production (right).}
    \label{fig:feynman}
\end{figure}

The signal process is dark photon bremsstrahlung followed by $A'\to e^+e^-$, as shown in figure~\ref{fig:feynman}. The main prompt background is $e^+e^-$ production through either real photon conversion or a process mediated by a virtual photon~\cite{Tsai:1974fa}, also shown in figure~\ref{fig:feynman}. Finite vertex resolution and multiple scattering can cause prompt vertices to be reconstructed at displaced positions.

Photonuclear and electronuclear reactions in the target provide an additional source of background. These reactions can produce $K^0_S$ mesons whose decays to two charged particles form genuine displaced vertices. When the daughter tracks are reconstructed under the electron and positron hypotheses, such decays can mimic the signal topology, although their calorimeter response and reconstructed invariant mass differ from those of the signal.

\section{The DarkSHINE Experiment}
\label{sec:detector}

\begin{figure}
    \centering
    \includegraphics[width=\linewidth]{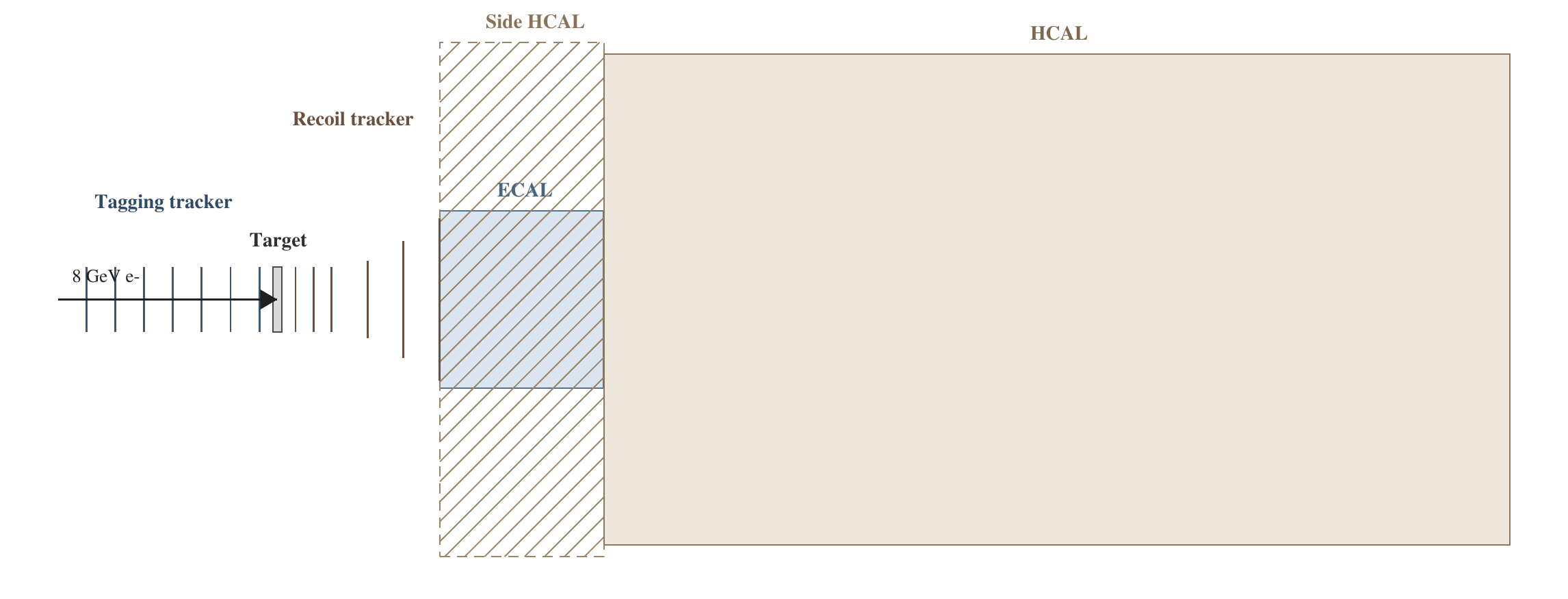}
    \caption{Schematic layout of the DarkSHINE detector for the visible decay search, including the tagging tracker, target, close tracker, recoil tracker, ECAL, and HCAL.}
    \label{fig:detector_schematic}
\end{figure}

DarkSHINE uses a continuous 8~GeV electron beam from the SHINE facility at an average rate of 10~MHz~\cite{chen2023prospective}. One year of operation corresponds to approximately $3\times10^{14}$ electrons on target (EOT). The beam strikes a tungsten target with a thickness of 0.1 radiation lengths ($0.33~\mathrm{mm}$). The detector comprises a silicon tracking system, an electromagnetic calorimeter (ECAL), and a hadronic calorimeter (HCAL), as shown in figure~\ref{fig:detector_schematic}~\cite{chen2023prospective}. The displaced vertex is reconstructed with the tracking system, while the calorimeter information is used to reject hadronic backgrounds.

The tracking system contains seven tagging layers upstream of the target and six recoil layers downstream. Each layer uses silicon strip sensors with two readout directions~\cite{Liu:2024DarkSHINEACLGAD}. Table~\ref{tab:tracker_config} summarizes the main configuration parameters. A uniform 0.3~T magnetic field is applied throughout the tracking volume. Relative to the design for the invisible decay search~\cite{chen2023prospective}, the recoil region is extended to improve the vertex resolution, and the magnetic field is reduced to increase acceptance for tracks with low momentum.

\begin{table}[htbp]
    \centering
    \begin{tabular}{ll}
        \toprule
        \textbf{Parameter} & \textbf{Value} \\
        \midrule
        Strip width & $30\,\mu\mathrm{m}$ \\
        Sensor thickness & $150\,\mu\mathrm{m}$ \\
        Tagging tracker positions & $-5,\,-13,\,-21,\,-29,\,-37,\,-45,\,-53~\mathrm{cm}$ \\
        Recoil tracker positions & $5,\,10,\,15,\,25,\,35,\,45~\mathrm{cm}$ \\
        Rotation angles & $(0^\circ,\,90^\circ)$ and $(-45^\circ,\,45^\circ)$ \\
        \bottomrule
    \end{tabular}
    \caption{Key parameters of the tracker configuration used in this analysis.}
    \label{tab:tracker_config}
\end{table}

To suppress prompt backgrounds, an additional close tracker layer is placed 2~mm downstream of the target. For a signal decay downstream of this layer, only the recoil electron crosses the close tracker and produces one hit. Prompt pair production can instead send two electrons and one positron through the layer. The hit multiplicity and deposited energy therefore help distinguish signal events from prompt backgrounds.

As a reference for the single particle signal, a minimum ionizing particle (MIP) has a mean stopping power of approximately $3.88~\mathrm{MeV/cm}$ in silicon~\cite{PDG:Silicon}, corresponding to a mean energy loss of $0.116~\mathrm{MeV}$ at normal incidence through the combined $300~\mu\mathrm{m}$ silicon thickness of one layer. We use the energy deposits recorded by \textsc{Geant4} and apply an additional $4~\mathrm{keV}$ smearing to model electronics noise. This corresponds approximately to an equivalent noise charge of $1000$ electrons, a representative scale for silicon strip readout~\cite{CMS:2009strip}.

\begin{figure}[htbp]
    \centering
    \includegraphics[width=0.8\linewidth]{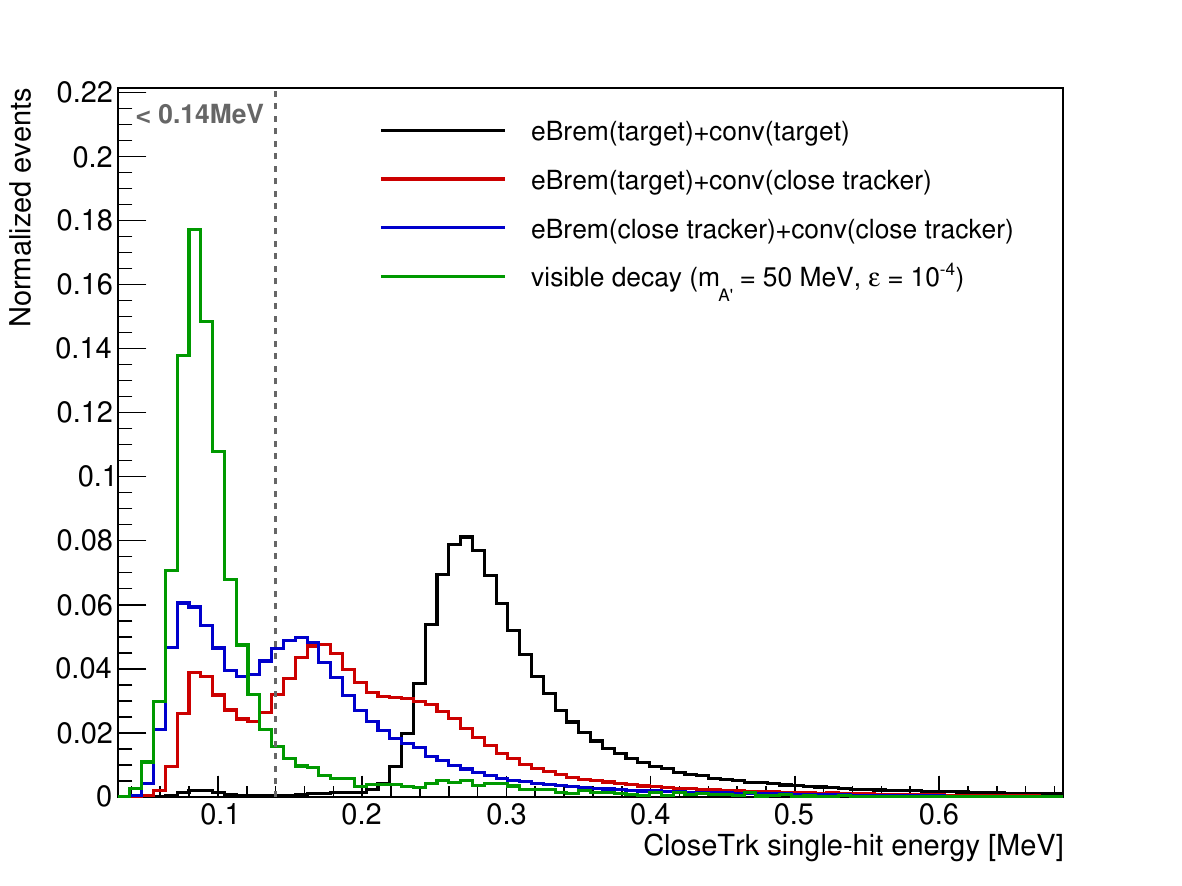}
    \caption{Normalized close tracker hit energy distributions for signal with $m_{A'}=50~\mathrm{MeV}$ and $\epsilon=10^{-4}$ (green) and backgrounds from electron bremsstrahlung followed by photon conversion. The black, red, and blue curves correspond to bremsstrahlung and conversion both in the target, bremsstrahlung in the target followed by conversion in the close tracker, and both processes in the close tracker, respectively. The gray dashed line marks the $0.14~\mathrm{MeV}$ threshold; hits with energies below this value are retained.}
    \label{fig:close_tracker_energy}
\end{figure}

Figure~\ref{fig:close_tracker_energy} compares the close tracker hit energy distributions for signal and photon conversion backgrounds. The signal distribution peaks below $0.14~\mathrm{MeV}$, whereas photon conversion in the target produces a peak at higher energy. Backgrounds involving conversion in the close tracker have broader distributions that overlap with the signal. We therefore require exactly one hit with smeared deposited energy below $0.14~\mathrm{MeV}$ to retain the main signal peak while suppressing backgrounds with larger energy deposits. Together, these requirements retain $63\%$ of the signal while rejecting most prompt backgrounds.

The dominant surviving background is electron bremsstrahlung in the target followed by photon conversion in the close tracker, as shown in Table~\ref{tab:close_tracker_statistics}. Including direct pair production, the total expected background decreases from approximately $6.8\times10^{12}$ to $8.0\times10^{10}$ events, a reduction factor of about 85.

\begin{table}[p]
    \centering
    \scriptsize
    \setlength{\tabcolsep}{4pt}
    \renewcommand{\arraystretch}{1.2}

    \subfloat[Electron bremsstrahlung followed by photon conversion.]{\begin{minipage}{0.98\linewidth}
    \centering
    \resizebox{\linewidth}{!}{%
    \begin{tabular}{lcccc}
        \toprule
        \textbf{Quantity}
        & \textbf{Visible decay}
        & \makecell{\textbf{eBrem (target)}\\
                    \textbf{+ conv (target)}}
        & \makecell{\textbf{eBrem (target)}\\
                    \textbf{+ conv (close tracker)}}
        & \makecell{\textbf{eBrem (close tracker)}\\
                    \textbf{+ conv (close tracker)}} \\
        \midrule

        \makecell[l]{Number of events in $3\times10^{14}$ EOT}
        & Not applicable
        & $\mathbf{5.9\times10^{12}}$
        & $3.4\times10^{11}$
        & $6.9\times10^{9}$ \\

        Close tracker hit number $=1$
        & $75\%$
        & $1.4\times10^{12}\;(24\%)$
        & $2.7\times10^{11}\;(79\%)$
        & $5.3\times10^{9}\;(76\%)$ \\

        Close tracker hit energy $<0.14\,\mathrm{MeV}$
        & $63\%$
        & $1.8\times10^{9}\;(0.03\%)$
        & $\mathbf{7.0\times10^{10}\;(20\%)}$
        & $2.5\times10^{9}\;(36\%)$ \\

        \bottomrule
    \end{tabular}
    }
    \end{minipage}}

    \par\medskip

    \subfloat[Direct electron and positron pair production.]{\begin{minipage}{0.98\linewidth}
    \centering
    \begin{tabular}{lccc}
        \toprule
        \textbf{Quantity}
        & \textbf{Visible decay}
        & \makecell{\textbf{ePairProd}\\
                    \textbf{(target)}}
        & \makecell{\textbf{ePairProd}\\
                    \textbf{(close tracker)}} \\
        \midrule

        \makecell[l]{Number of events in $3\times10^{14}$ EOT}
        & Not applicable
        & $5.4\times10^{11}$
        & $2.4\times10^{10}$ \\

        Close tracker hit number $=1$
        & $75\%$
        & $9.0\times10^{10}\;(17\%)$
        & $1.8\times10^{10}\;(76\%)$ \\

        Close tracker hit energy $<0.14\,\mathrm{MeV}$
        & $63\%$
        & $6.0\times10^{8}\;(0.1\%)$
        & $5.2\times10^{9}\;(22\%)$ \\

        \bottomrule
    \end{tabular}
    \end{minipage}}

    \caption{Signal efficiencies and expected background yields after the close tracker selections for $m_{A'}=50~\mathrm{MeV}$ and $\epsilon=10^{-4}$. Background yields correspond to $3\times10^{14}$ EOT, and values in parentheses are selection efficiencies. Here eBrem denotes electron bremsstrahlung, conv denotes photon conversion, and ePairProd denotes direct electron and positron pair production.}
    \label{tab:close_tracker_statistics}
\end{table}
\section{Simulation and Event Reconstruction}
\label{sec:sim_and_recon}

Signal production cross sections and event kinematics are calculated at tree level with \textsc{CalcHEP}~\cite{Belyaev:2012qa}, and the generated events are propagated through a detailed \textsc{Geant4} detector simulation~\cite{Agostinelli:2002hh}. We generate $10^4$ events for each dark photon mass and coupling point. The prompt-background sample contains approximately $3.7\times10^8$ events in which electron bremsstrahlung in the target is followed by photon conversion to $e^+e^-$ in the close tracker. Dedicated photonuclear and electronuclear samples, containing approximately $10^7$ simulated events in total, are generated with \textsc{Geant4} cross-section biasing and processed through the same simulation and reconstruction chain.

Reliable track reconstruction is essential for identifying displaced vertices. Although the continuous beam produces little pileup, the forward event topology and multiple scattering of particles with low momentum make track finding challenging. We therefore develop TrackNet, an updated version of the GNN track finding algorithm introduced in Ref.~\cite{Lu:2024gnn}. TrackNet adds Convolutional Block Attention Modules (CBAMs)~\cite{Woo:2018cbam} to the node and edge embedding branches and uses a gated fusion module to update their representations.

\begin{figure}[htbp]
    \centering
    \includegraphics[width=0.60\linewidth]{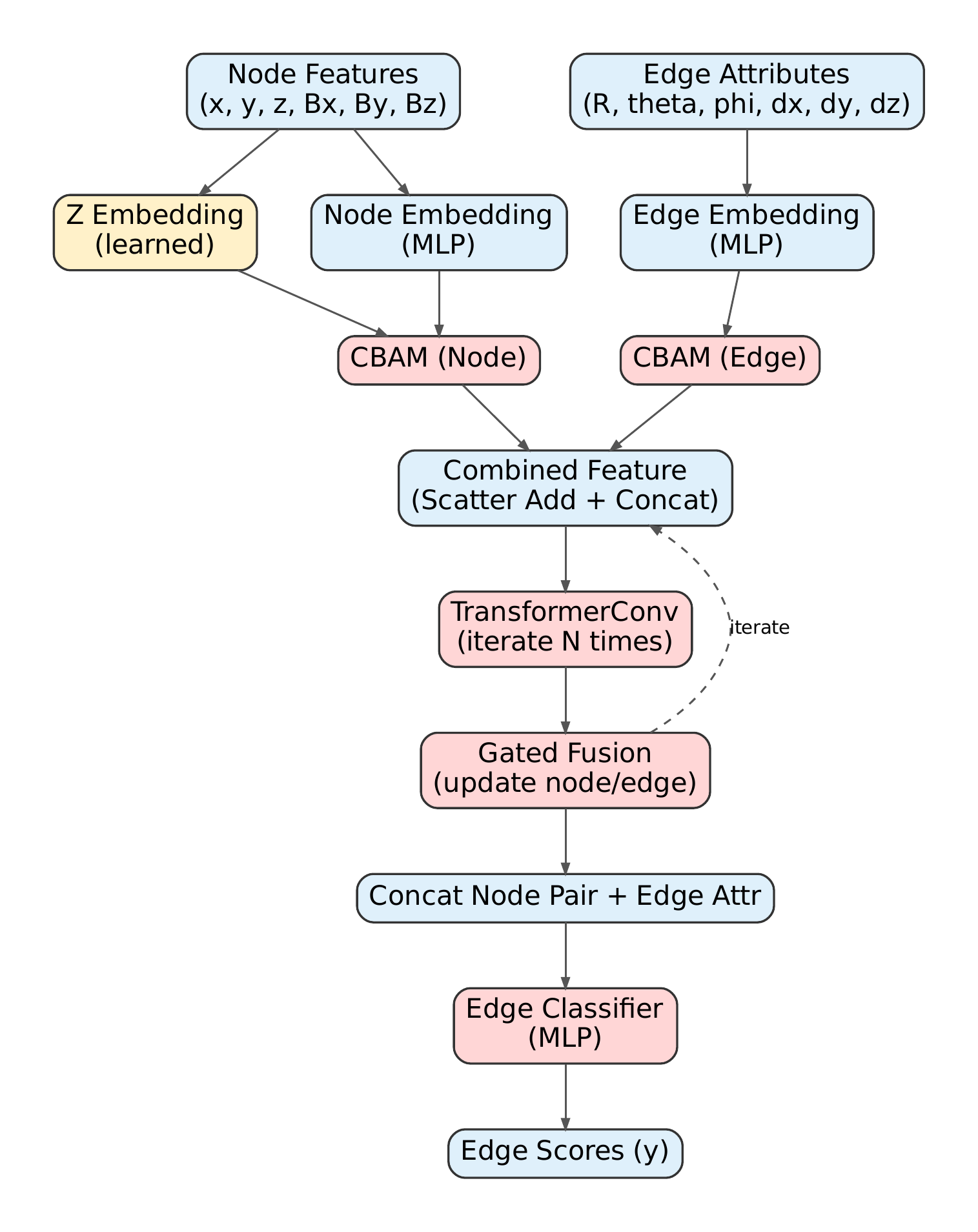}
    \caption{TrackNet architecture. CBAMs refine the node and edge embeddings, a gated fusion module updates both representations, and the final classifier assigns a score to each candidate hit pair for the global track search.}
    \label{fig:trackNet}
\end{figure}

\begin{figure}[htbp]
    \centering
    \includegraphics[width=0.8\linewidth]{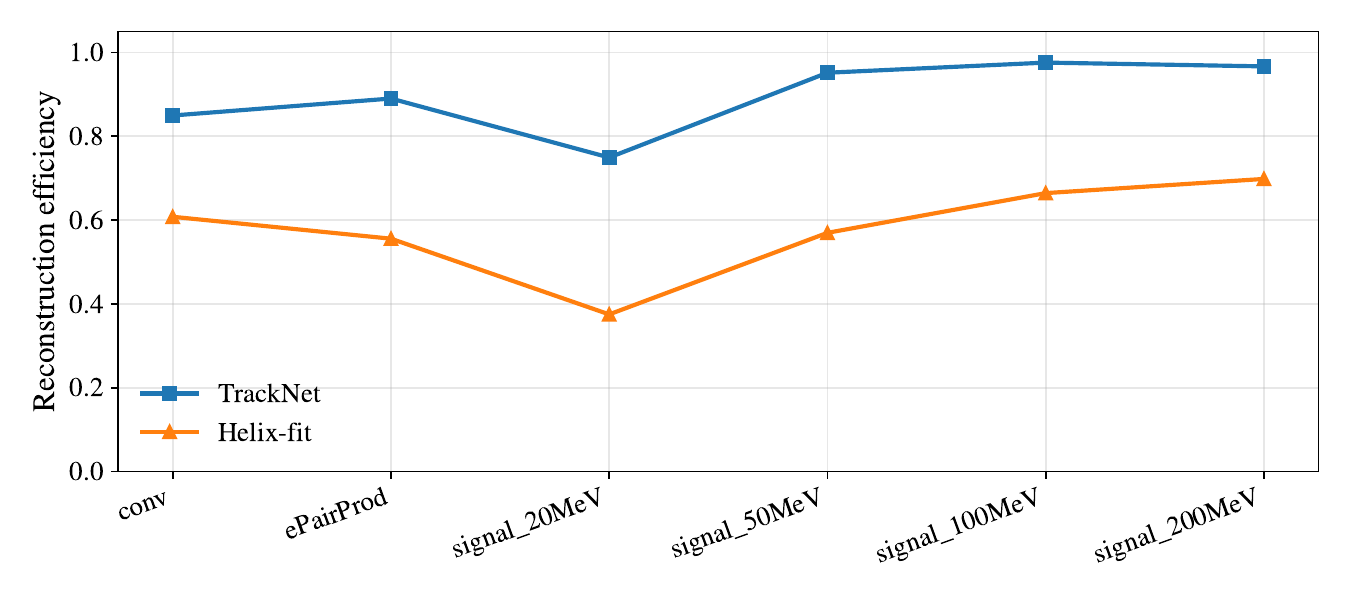}
    \caption{Reconstruction efficiency of TrackNet and conventional helix fitting for photon conversion and direct pair production backgrounds and signal samples at several dark photon masses. The efficiency is defined as the number of correctly recovered hits divided by the total number of true hits belonging to the tracks to be reconstructed.}
    \label{fig:track_eff}
\end{figure}

TrackNet takes hit coordinates and local magnetic field components as input. It assigns each candidate hit pair a score representing the probability that both hits belong to the same track. Candidate edges connect only adjacent tracker layers. Dijkstra's algorithm~\cite{Dijkstra:1959} then identifies the track candidates with the highest total scores. Minimum edge score and shared hit requirements suppress spurious combinations. Figure~\ref{fig:trackNet} shows the network architecture. We define the reconstruction efficiency as the fraction of true track hits correctly recovered by the track finding algorithm. Relative to conventional helix fitting, TrackNet increases this efficiency from about 60\% to 90\%, as shown in figure~\ref{fig:track_eff}. The selected tracks are fitted with the Kalman filter implemented in \textsc{GenFit2}~\cite{Hoppner:2009rq,Bilka:2019genfit2}.

Vertices are reconstructed with \textsc{Rave}~\cite{Waltenberger:2007zz} through the \textsc{GenFit2} interface.\begin{NoHyper}\footnote{During this work, we identified and corrected a bug in the track propagator implementation in \textsc{GFRave}.}\end{NoHyper} Each oppositely charged track pair is fitted with a Kalman vertex fit. The incoming track from the tagging tracker is extrapolated to the target to estimate the interaction point, relative to which the outgoing track impact parameters are calculated.
\section{Displaced Vertex Analysis}
\label{sec:analysis}

Events must contain at least one vertex candidate with exactly two tracks. If an event contains multiple valid candidates, we select the one with the smallest $\chi^2$. We then apply the close tracker requirements described in section~\ref{sec:detector}: exactly one hit near the interaction point with deposited energy below $0.14~\mathrm{MeV}$. Candidates with shared hits are rejected. We also require a total ECAL energy above $7.5~\mathrm{GeV}$ and a total HCAL energy below $50~\mathrm{MeV}$ to reject photonuclear and electronuclear events.

\begin{figure}[htbp]
    \centering
    \includegraphics[width=0.62\linewidth]{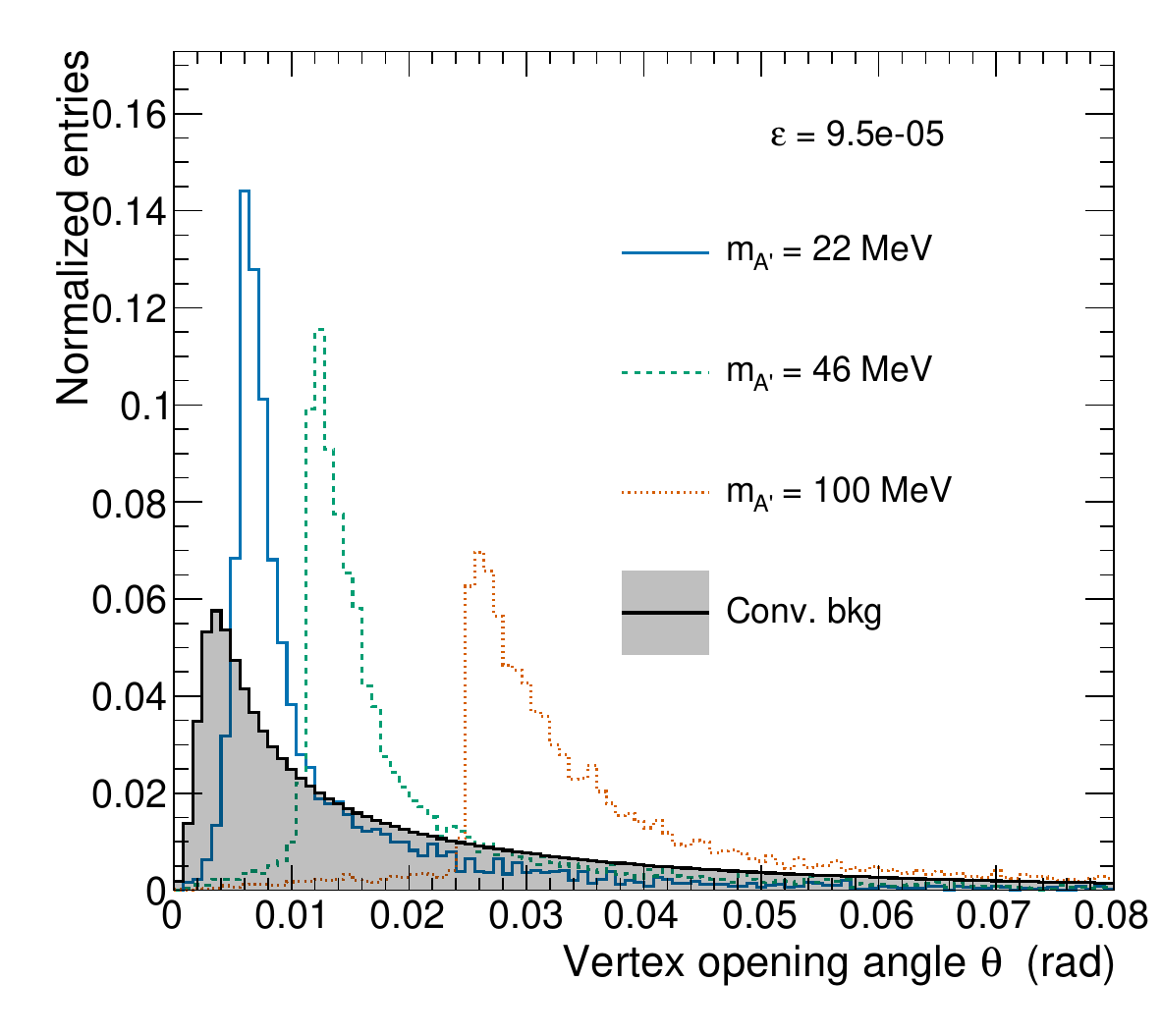}
    \caption{Normalized vertex opening angle distributions for signal samples with $m_{A'}=22$, 46, and $100~\mathrm{MeV}$ at $\epsilon=9.5\times10^{-5}$ and for the photon conversion background.}
    \label{fig:vertex_theta}
\end{figure}

The forward event topology allows poorly reconstructed prompt vertices to populate a broad region of the tracker. Baseline selections are applied to the vertex fit quality, minimum track momentum, vertex dispersion, invariant mass, and opening angle. The opening angle is particularly important because nearly parallel tracks constrain the vertex position along $z$ poorly. The nonzero dark photon mass gives signal events larger opening angles than photon conversion events, as shown in figure~\ref{fig:vertex_theta}. We require $\theta_{\mathrm{vtx}}>0.012~\mathrm{rad}$ and impose an additional condition based on the invariant mass:
\begin{equation}
\theta_{\mathrm{vtx}} > 0.000237\left(\frac{m_{\mathrm{inv}}}{\mathrm{MeV}/c^2}\right)+0.00058.
\label{eq:mass_theta_cut}
\end{equation}

\begin{figure}[htbp]
    \centering
    \includegraphics[width=0.70\linewidth]{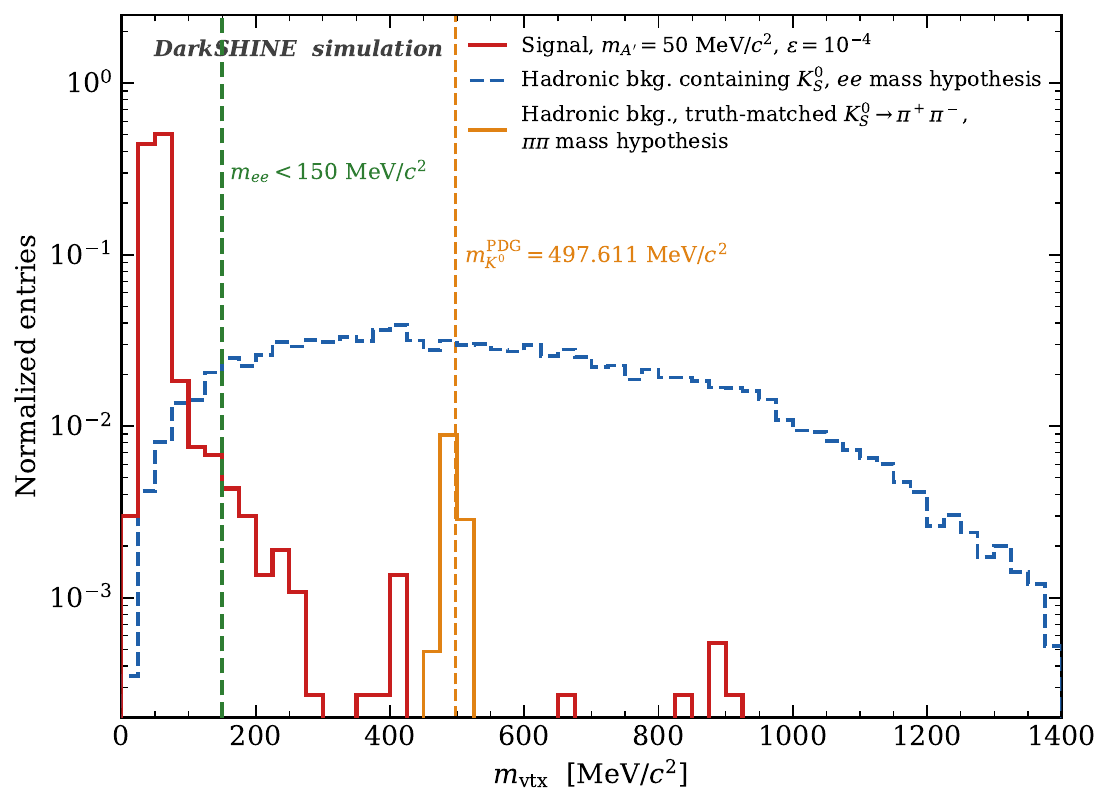}
    \caption{Normalized vertex invariant-mass distributions for the benchmark signal (red) and hadronic background containing a target-produced $K^0_S$, reconstructed under the $e^+e^-$ hypothesis (blue). The orange distribution shows truth-matched $K^0_S\to\pi^+\pi^-$ decays under the pion-mass hypothesis. Dashed lines mark the $m_{ee}<150~\mathrm{MeV}/c^2$ requirement and the $K^0_S$ mass.}
    \label{fig:k0s_invmass}
\end{figure}

Unlike prompt pair-production backgrounds, decays of strange hadrons form genuine displaced vertices. Their charged daughters are reconstructed as electrons, resulting in an invariant mass that is generally larger than the signal mass range. Figure~\ref{fig:k0s_invmass} compares this distribution with the benchmark signal. We therefore impose the additional requirement
\begin{equation}
20 < m_{\mathrm{inv}} < 150~\mathrm{MeV}/c^2.
\label{eq:mass_window}
\end{equation}
The upper bound retains $98.5\%$ of the benchmark signal at this stage while rejecting $94\%$ of the hadronic background, with no signal loss after the full selection.

\begin{figure}[htbp]
    \centering
    \includegraphics[width=0.72\linewidth]{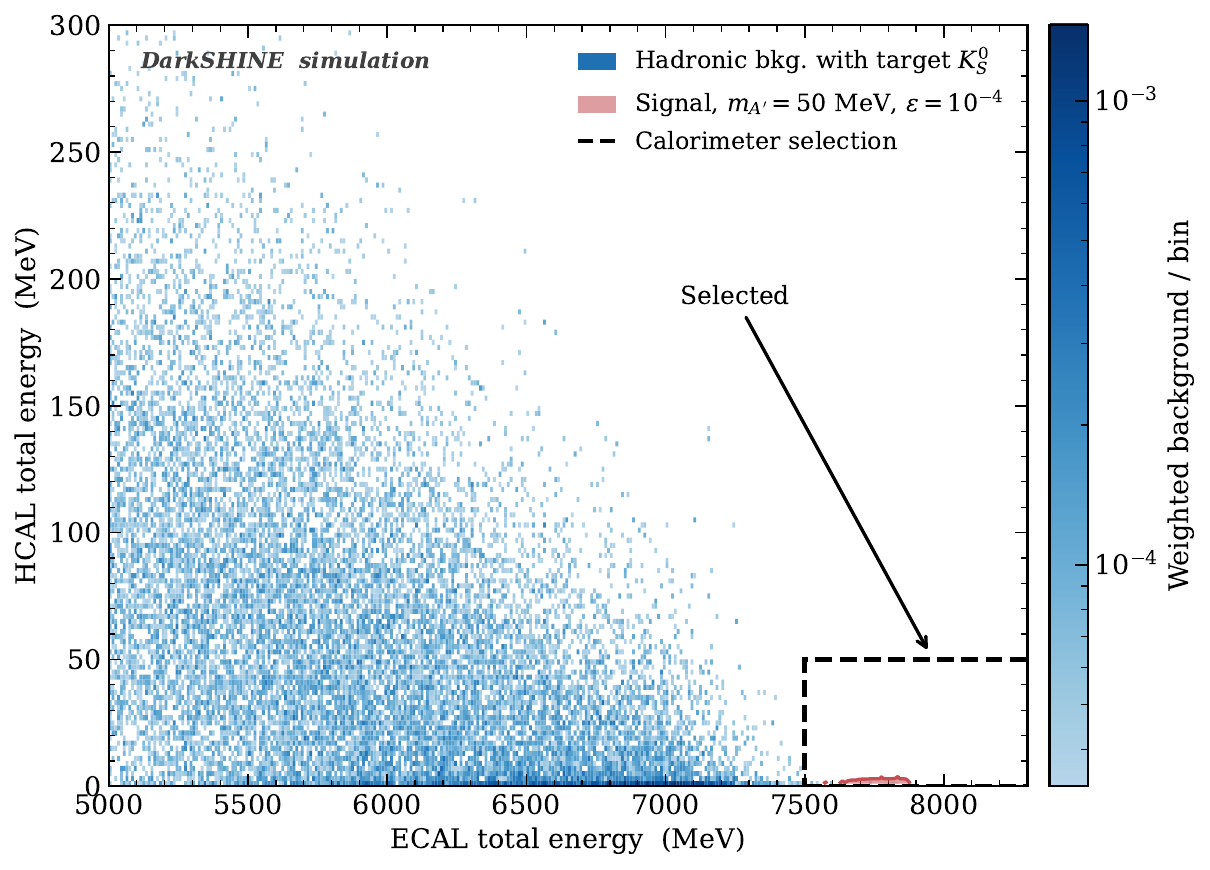}
    \caption{Total energy deposited in the ECAL and HCAL for hadronic events with a $K^0_S$ produced in the target (blue) and signal with $m_{A'}=50~\mathrm{MeV}$ and $\epsilon=10^{-4}$ (red contours). The dashed box indicates the selected region, $E_{\mathrm{ECAL}}>7.5~\mathrm{GeV}$ and $E_{\mathrm{HCAL}}<50~\mathrm{MeV}$.}
    \label{fig:k0s_calo}
\end{figure}

Hadronic reactions also carry part of the beam energy into neutral hadrons and strange baryons, reducing the ECAL energy and, in some events, producing an HCAL deposit. As shown in figure~\ref{fig:k0s_calo}, the calorimeter requirements retain $99.4\%$ of the signal averaged over the simulated mass points, while accepting only approximately $10^{-4}$ of the hadronic events with a reconstructed vertex.

\begin{figure}[htbp]
    \centering
    \includegraphics[width=0.9\linewidth]{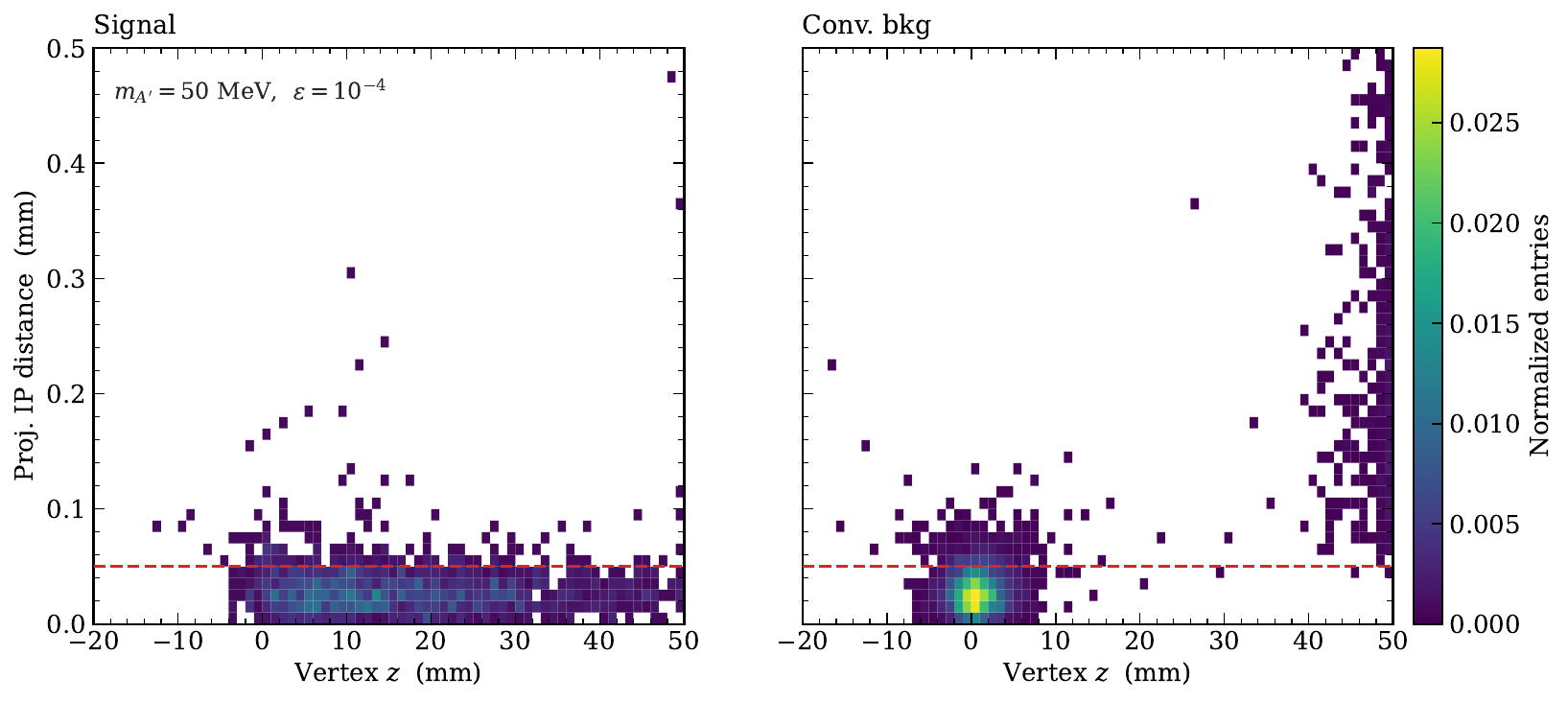}
    \caption{Normalized distributions of projected IP distance versus reconstructed vertex $z$ for signal with $m_{A'}=50~\mathrm{MeV}$ and $\epsilon=10^{-4}$ (left) and photon conversion background (right). The red dashed line marks $d_{\mathrm{projIP}}=0.05~\mathrm{mm}$; events below the line are retained.}
    \label{fig:projIPDist}
\end{figure}

Following the HPS analysis~\cite{Adrian:2023hps}, we use the consistency of the projected interaction point to suppress fake displaced vertices from multiple scattering. The reconstructed vertex momentum is extrapolated to the target and compared with the interaction point predicted by the tagging track. Their separation defines $d_{\mathrm{projIP}}$. As shown in figure~\ref{fig:projIPDist}, background vertices reconstructed at large $z$ often have large $d_{\mathrm{projIP}}$, whereas most signal events remain near the predicted interaction point. We require $d_{\mathrm{projIP}}<0.05~\mathrm{mm}$.

\begin{figure}[htbp]
    \centering
    \includegraphics[width=0.66\linewidth]{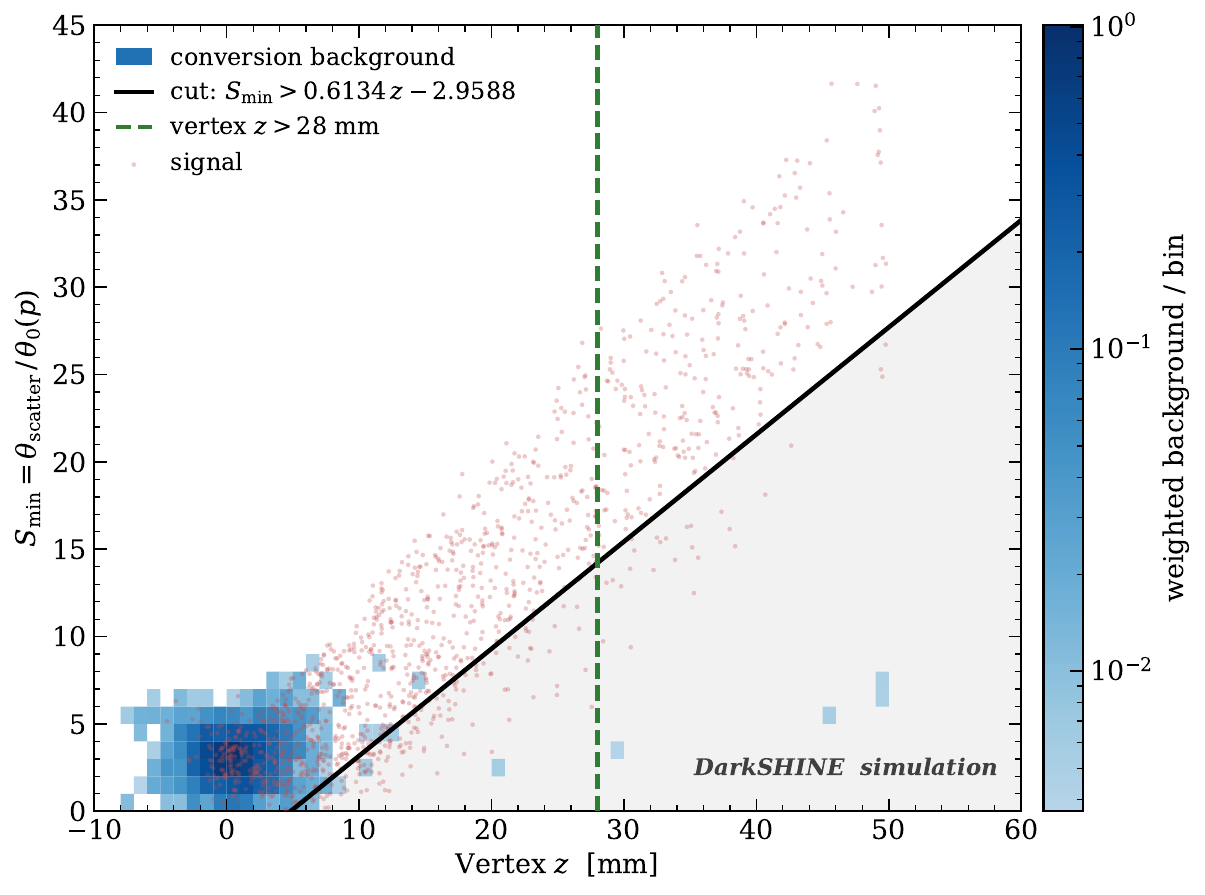}
    \caption{Minimum multiple scattering significance versus reconstructed vertex $z$ for photon conversion background (blue) and signal (red). The solid black line shows the selection in equation~\eqref{eq:smin_cut}. The dashed green line at $z=28~\mathrm{mm}$ marks the lower boundary of the signal region.}
    \label{fig:Smin}
\end{figure}

To reject vertices formed from tracks scattered in the first recoil layer, we define a multiple scattering significance that accounts for momentum. The RMS scattering angle is approximated by~\cite{Lynch1991}
\begin{equation}
\theta_0 =
\frac{13.6~\mathrm{MeV}}{\beta p}
\sqrt{\frac{L}{X_0}}
\left[
1+0.038\ln\left(\frac{L}{X_0}\right)
\right].
\label{eq:highland}
\end{equation}
Here $L=0.300~\mathrm{mm}$ is the combined silicon thickness of the two strip planes in one layer, $X_0=93.69~\mathrm{mm}$ is the radiation length of silicon, and $\beta p=p^2/\sqrt{p^2+m_e^2}$. For each track, the apparent scattering angle under the prompt origin hypothesis is $\theta_{\mathrm{scatter}}=r_{\mathrm{IP}}/z_{\mathrm{first}}$, where $r_{\mathrm{IP}}$ is the impact parameter magnitude and $z_{\mathrm{first}}=50.33~\mathrm{mm}$ is the distance from the interaction point to the first recoil layer. The significance is
\begin{equation}
S_i =
\frac{\theta_{\mathrm{scatter},i}}{\theta_0(p_i)}=
\frac{\theta_{\mathrm{scatter},i}\,\beta_i p_i}
{13.6~\mathrm{MeV}\sqrt{L/X_0}
\left[1+0.038\ln\left(L/X_0\right)\right]},
\label{eq:ms_significance}
\end{equation}
for each vertex track. Both tracks from a genuine displaced decay are inconsistent with the prompt interaction point, whereas a scattering kink or accidental pairing often leaves one track consistent with a prompt origin. We therefore define $S_{\min}=\min(S_0,S_1)$ and require
\begin{equation}
S_{\min}>0.6134\left(\frac{z}{\mathrm{mm}}\right)-2.9588.
\label{eq:smin_cut}
\end{equation}
This boundary follows the lower edge of the signal distribution in figure~\ref{fig:Smin}. Together with the vertex-quality, kinematic, calorimeter, and close tracker criteria described above, this requirement completes the event selection summarized in Table~\ref{tab:selection_cuts}.

\begin{table}[htbp]
\centering
\small
\begin{tabular}{p{0.38\linewidth}p{0.54\linewidth}}
\toprule
\textbf{Variable} & \textbf{Selection} \\
\midrule
Valid vertex candidate & Exactly two associated tracks; select the candidate with minimum $\chi^2_{\mathrm{vtx}}$ \\
Close tracker hit multiplicity & $N_{\mathrm{hit,near~IP}}=1$ \\
Close tracker hit energy & $E_{\mathrm{hit}}<0.14~\mathrm{MeV}$ \\
ECAL energy & $E_{\mathrm{ECAL}}>7.5~\mathrm{GeV}$ \\
HCAL energy & $E_{\mathrm{HCAL}}<50~\mathrm{MeV}$ \\
Shared hits & $N_{\mathrm{shared}}=0$ \\
Vertex fit quality & $\chi^2_{\mathrm{vtx}}<3$ \\
Minimum track momentum & $p_{\min} > 1000~\mathrm{MeV}/c$ \\
Vertex opening angle & $\theta_{\mathrm{vtx}} > 0.012~\mathrm{rad}$ \\
Vertex dispersion & $D_{\mathrm{vtx}} < 0.2~\mathrm{mm}$ \\
Invariant mass & $20<m_{\mathrm{inv}}<150~\mathrm{MeV}/c^2$ \\
Mass and angle correlation & $\theta_{\mathrm{vtx}}>0.000237(m_{\mathrm{inv}}/[\mathrm{MeV}/c^2])+0.00058$ \\
Projected IP distance & $d_{\mathrm{projIP}} < 0.05~\mathrm{mm}$ \\
Multiple scattering significance & $S_{\min}>0.6134(z/\mathrm{mm})-2.9588$ \\
Signal region for displaced vertices & $28<z<49~\mathrm{mm}$ \\
\bottomrule
\end{tabular}
\caption{Selections used in the displaced vertex analysis.}
\label{tab:selection_cuts}
\end{table}

\begin{figure}[htbp]
    \centering
    \includegraphics[width=0.66\linewidth]{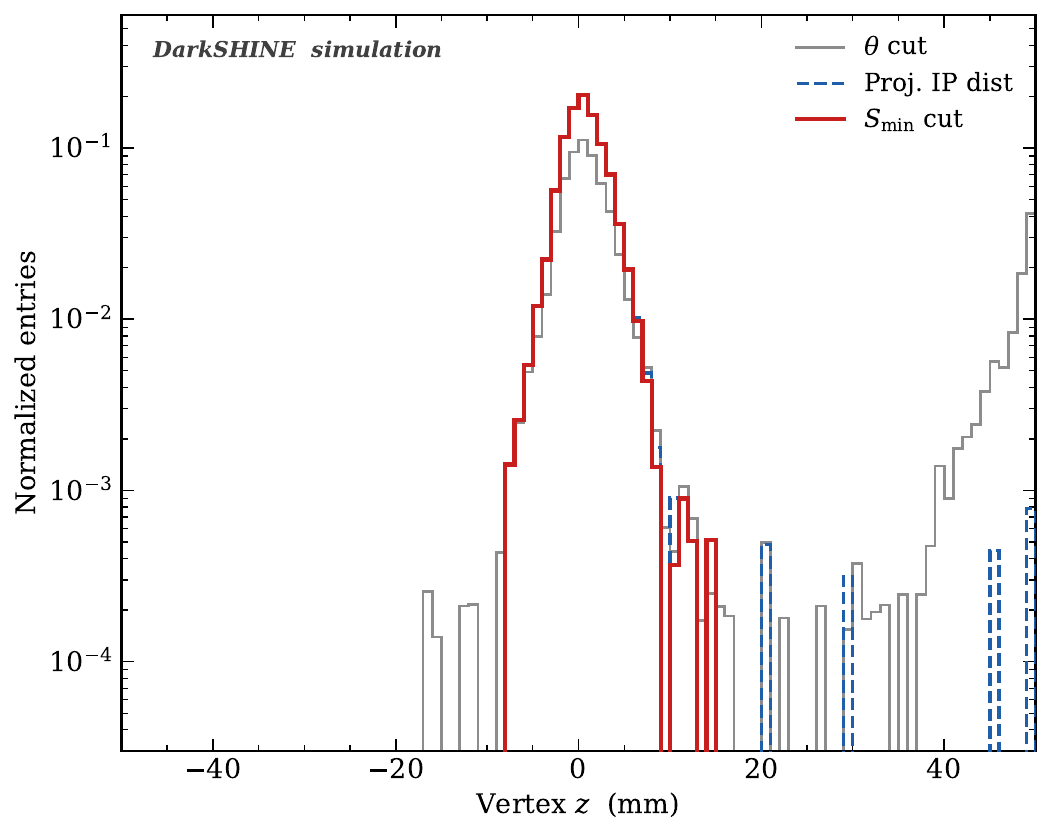}
    \caption{Normalized photon conversion background distributions in reconstructed vertex $z$ after the opening angle selection (gray), projected IP distance selection (blue dashed), and $S_{\min}$ selection (red).}
    \label{fig:vtx_z2}
\end{figure}

\begin{figure}[htbp]
    \centering
    \includegraphics[width=0.66\linewidth]{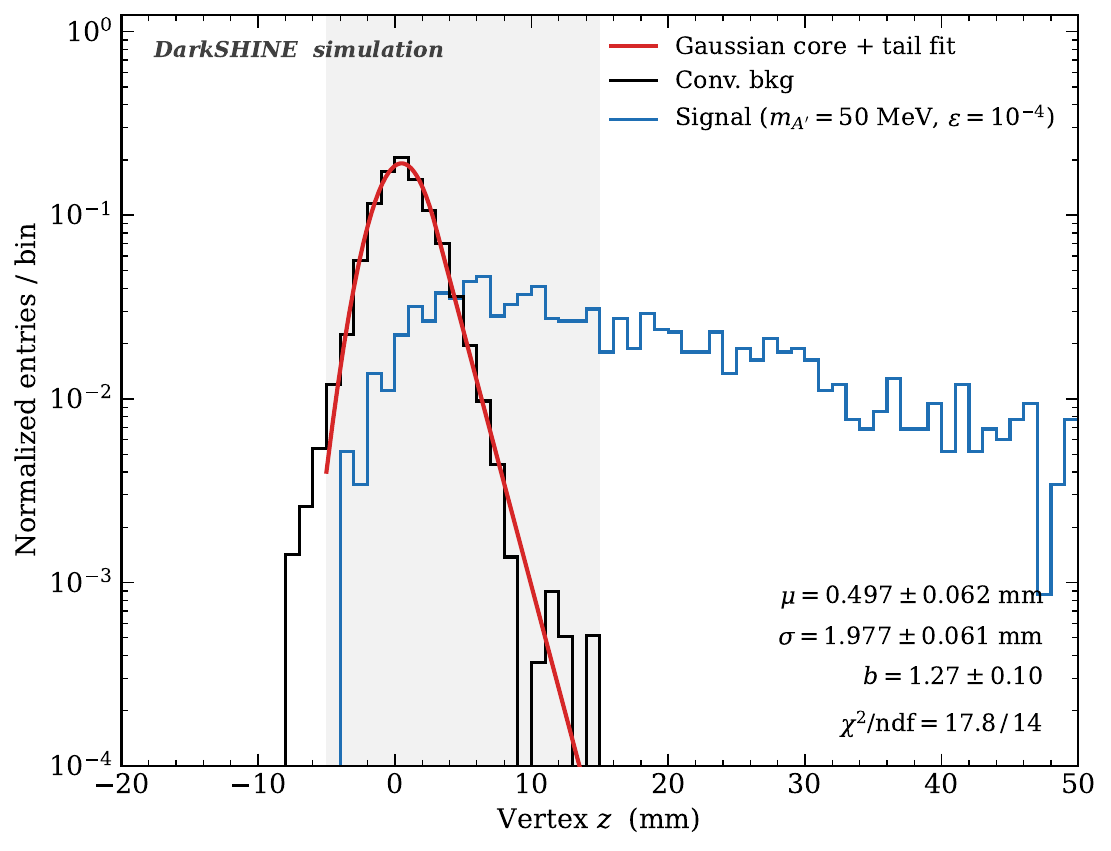}
    \caption{Normalized reconstructed vertex $z$ distributions after all selections except the final $z$ requirement for photon conversion background (black) and signal with $m_{A'}=50~\mathrm{MeV}$ and $\epsilon=10^{-4}$ (blue). The red curve is the background fit defined in equation~\eqref{eq:vertex_z_fit} within the shaded range.}
    \label{fig:fit_vertex_z}
\end{figure}

Figure~\ref{fig:vtx_z2} shows how the projected interaction-point and multiple-scattering requirements suppress the long tail of the photon conversion background. After all selections except the final $z$ requirement, the remaining photon conversion background is concentrated near the target, as shown in figure~\ref{fig:fit_vertex_z}. Following the HPS displaced vertex analysis~\cite{Adrian:2023hps}, we model the distribution with a Gaussian core joined continuously to an exponential tail:
\begin{equation}
F(z)=
\begin{cases}
A\exp\!\left[-\dfrac{(z-\mu_z)^2}{2\sigma_z^2}\right],
& \dfrac{z-\mu_z}{\sigma_z}<b, \\[8pt]
A\exp\!\left[\dfrac{b^2}{2}-b\dfrac{z-\mu_z}{\sigma_z}\right],
& \dfrac{z-\mu_z}{\sigma_z}\geq b.
\end{cases}
\label{eq:vertex_z_fit}
\end{equation}
Here $A$ is the normalization, $\mu_z$ and $\sigma_z$ describe the Gaussian core, and $b$ sets the transition to the exponential tail and its slope. Integrating the fitted tail predicts 0.1 background events per $3\times10^{14}$~EOT above $z=28~\mathrm{mm}$. The signal region is therefore defined as $28<z<49~\mathrm{mm}$, ending before the first recoil layer.

Before the calorimeter and invariant-mass requirements are applied, strange-hadron decays can contribute hundreds of events in the signal region. No photonuclear or electronuclear simulated event survives the complete selection. Because of the limited sample size, the efficiencies of the calorimeter, close tracker, and vertex requirements are evaluated independently for events containing a reconstructed vertex and combined assuming factorization of these requirements. The resulting 90\% confidence-level upper limits are summarized in Table~\ref{tab:hadronic_eff}. The combined hadronic background is below $4.6\times10^{-3}$ events per $3\times10^{14}$ EOT, well below the prompt-background prediction of 0.1 events.

\begin{table}[htbp]
\centering
\small
\begin{tabular}{lccc}
\toprule
 & \makecell{Signal\\$m_{A'}=50~\mathrm{MeV}$} & Photonuclear & Electronuclear \\
\midrule
Calorimeter requirements & $99.5\%$ & $1.1\times10^{-4}$ & $2.0\times10^{-4}$ \\
Close tracker requirements & $62.8\%$ & $3.6\times10^{-2}$ & $4.5\times10^{-2}$ \\
Vertex requirements & $3.6\%$ & $<1.3\times10^{-4}$ & $<3.2\times10^{-4}$ \\
\midrule
Combined efficiency & $2.2\%$ & $<4.9\times10^{-10}$ & $<2.8\times10^{-9}$ \\
Events in $3\times10^{14}$ EOT & --- & $<1.3\times10^{-3}$ & $<3.3\times10^{-3}$ \\
\bottomrule
\end{tabular}
\caption{Efficiencies of the calorimeter, close tracker, and vertex requirements, evaluated independently for events with a reconstructed vertex. The signal benchmark has $m_{A'}=50~\mathrm{MeV}$ and $\epsilon=10^{-4}$, and the hadronic samples require a $K^0_S$ produced in the target. Background yields are obtained from the 90\% confidence-level efficiency bounds under the factorization approximation.}
\label{tab:hadronic_eff}
\end{table}

\begin{figure}[htbp]
    \centering
    \includegraphics[width=0.72\linewidth]{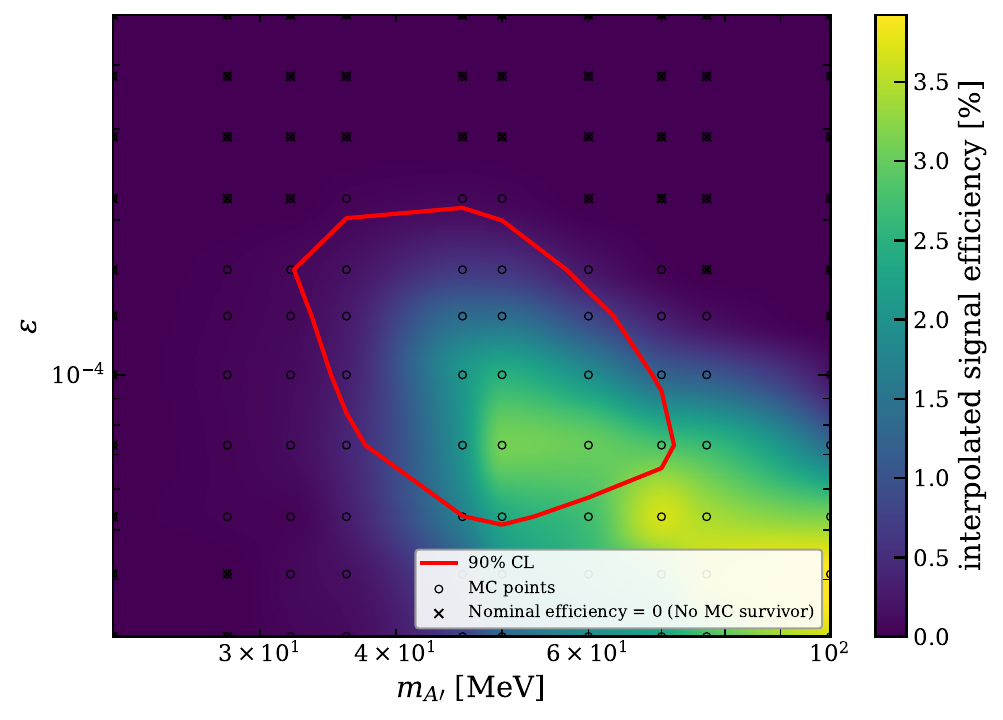}
    \caption{Interpolated signal efficiency after the full event selection as a function of dark photon mass and kinetic mixing. Open circles indicate simulated parameter points, crosses indicate points with no surviving simulated event, and the red curve shows the projected sensitivity contour at 90\% confidence level for three years of DarkSHINE operation, corresponding to $9\times10^{14}$~EOT and 0.3 expected background events.}
    \label{fig:signal_efficiency}
\end{figure}

We apply the full selection to signal samples generated over a grid of dark photon masses and couplings. Figure~\ref{fig:signal_efficiency} shows the efficiency evaluated at the simulated points and interpolated across the parameter space. The efficiency reaches approximately $3.8\%$ in the most favorable region. The calorimeter and invariant-mass requirements reduce the signal efficiency by less than $1\%$ on average and leave the exclusion contour unchanged. Crosses mark simulated points at which no event survives the full selection, for which the nominal efficiency is set to zero. The red curve shows the projected sensitivity contour at 90\% confidence level for three years of data taking, assuming 0.3 expected background events in the signal region. The hadronic contribution is negligible at the quoted precision.

\begin{figure}[tbp]
    \centering
    \includegraphics[width=0.78\linewidth]{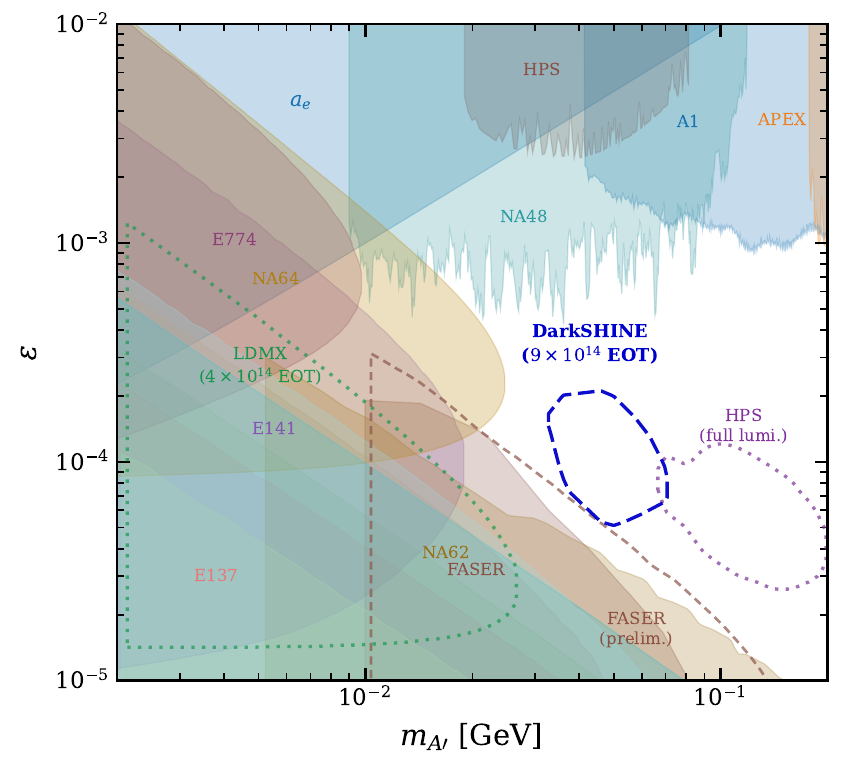}
    \caption{Median expected DarkSHINE exclusion sensitivity at 90\% confidence level for three years of data taking, assuming 0.3 expected background events. The Poisson $\mathrm{CL}_s$ contour corresponds to 2.30 selected signal events. Existing exclusion limits are compiled with \textnormal{\textsc{DarkCast}}~\cite{Ilten:2018crw}. The brown curve shows the preliminary FASER result~\cite{FASER:2026conf}. The purple and green dotted curves show the projected sensitivities of HPS for its full planned integrated luminosity~\cite{Baltzell:2022rpd} and LDMX for $4\times10^{14}$~EOT~\cite{LDMX:2026mrx}, respectively.}
    \label{fig:limit}
\end{figure}

Figure~\ref{fig:limit} shows the median expected exclusion sensitivity at 90\% confidence level for three years of data taking, corresponding to $9\times10^{14}$ EOT and 0.3 expected background events. We use an exact Poisson $\mathrm{CL}_s$ construction~\cite{Read:2002hq} with fixed expected yields and no systematic uncertainties. The median count for this background expectation is zero, giving $\mathrm{CL}_s=e^{-s}$, where $s$ is the expected signal yield after all selections; the contour therefore corresponds to $s=-\ln(0.10)\simeq2.30$ events. Existing limits are compiled with \textsc{DarkCast}~\cite{Ilten:2018crw} from references~\cite{Merkel:2014avp,Abrahamyan:2011gv,Tsai:2019mtm,Andreas:2012mt,Riordan:1987aw,Bross:1989mp,Bjorken:1988as,Adrian:2018scb,Konaka:1986cb,Batley:2015lha,Banerjee:2019hmi,Astier:2001ck,Davier:1989wz,Bodas:2021fsy,NA62:2023nhs,FASER:2023tle}.

For comparison, we also show the projected sensitivities of the HPS displaced vertex search for its full planned integrated luminosity~\cite{Baltzell:2022rpd} and the LDMX visible decay search for $4\times10^{14}$~EOT~\cite{LDMX:2026mrx}. These projections are shown separately from the existing exclusion limits.

\section{Conclusion}
\label{sec:conclusion}

We present a full simulation study of visible dark photon decays at DarkSHINE. Signal production cross sections and event kinematics are calculated at tree level using \textsc{CalcHEP}, and the generated events are propagated through a detailed \textsc{Geant4} detector model. The analysis combines the close tracker, calorimeter information, TrackNet reconstruction, Kalman vertex fitting, and selections on vertex and track observables to suppress prompt backgrounds dominated by photon conversion, as well as displaced strange-hadron decays from photonuclear and electronuclear reactions.

The projected IP distance and the multiple scattering significance $S_{\min}$, which accounts for momentum, efficiently reject fake displaced vertices caused by scattering in the tracker. Calorimeter and invariant-mass requirements reduce the hadronic contribution below $4.6\times10^{-3}$ events per year with a negligible loss of signal efficiency. A fit to the remaining prompt background predicts 0.1 events per year in the signal region $28<z<49~\mathrm{mm}$. For three years of data taking, corresponding to $9\times10^{14}$ electrons on target, we project the sensitivity at 90\% confidence level assuming 0.3 expected background events. These results demonstrate that visible decays at DarkSHINE can provide a complementary probe of dark photons below the GeV scale.

\section*{Data Availability Statement}
Selected data products generated in this study, including the signal efficiencies on the simulated mass--coupling grid, the dark photon production cross sections, and the exclusion contour of figure~\ref{fig:limit}, are available at \url{https://doi.org/10.5281/zenodo.23093250}.

\section*{Code Availability Statement}
The code that derives the exclusion contour from the signal efficiencies is available at \url{https://doi.org/10.5281/zenodo.23093250}.

\acknowledgments
The authors thank Shao-Feng Ge for his assistance with the signal sample generation and for helpful comments on the theoretical aspects of the analysis. This work was supported by the National Key R\&D Program of China (Grant Nos. 2024YFA1610600, 2024YFA1610603, 2023YFA1606904, and 2023YFA1606900), the National Natural Science Foundation of China (Grant Nos. 12475108 and 12150006), and the Shanghai Pilot Program for Basic Research---Shanghai Jiao Tong University (Grant No. 21TQ1400209). Zejia Lu is supported by the T. D. Lee Scholarship.

\bibliographystyle{JHEP}
\bibliography{references}

\end{document}